\documentclass[conference]{IEEEtran}
  
\usepackage{cite}
\usepackage{balance}

  \usepackage[pdftex]{graphics} 
\ifCLASSINFOpdf

\else
 
\fi

\usepackage{array}
\usepackage[font=footnotesize,caption=false]{subfig} 
\usepackage{stfloats}
\usepackage{url}
  \usepackage{color}

\begin{document}

\IEEEoverridecommandlockouts
\IEEEpubid{\makebox[\columnwidth]{~\\~\copyright~2015 IEEE~~~~\\
Preprint \hfill} \hspace{\columnsep}\makebox[\columnwidth]{ }}
 
\title{Chiminey: Reliable Computing and  Data Management  Platform in the Cloud}

\author{
\IEEEauthorblockN{
Iman I. Yusuf\IEEEauthorrefmark{1},
Ian E. Thomas\IEEEauthorrefmark{2},
Maria Spichkova\IEEEauthorrefmark{2},
Steve Androulakis\IEEEauthorrefmark{3},
Grischa  R. Meyer\IEEEauthorrefmark{3},\\
Daniel W. Drumm\IEEEauthorrefmark{2},
George Opletal\IEEEauthorrefmark{2},
Salvy P. Russo\IEEEauthorrefmark{2}
Ashley M.  Buckle\IEEEauthorrefmark{3} and
Heinz W. Schmidt\IEEEauthorrefmark{2}
}
\IEEEauthorblockA{\IEEEauthorrefmark{1}Applied Data Science, 
iyusuf@appds.com.au
}
\IEEEauthorblockA{\IEEEauthorrefmark{2} RMIT University, Australia \\ 
\{ian.edward.thomas, maria.spichkova, daniel.drumm, george.opletal, salvy.russo, heinz.schmidt\}@rmit.edu.au
}

\IEEEauthorblockA{\IEEEauthorrefmark{3} 
Monash University,  Australia\\
 \{steve.androulakis, grischa.meyer, ashley.buckle\}@monash.edu
}
 
}

\maketitle

\begin{abstract}

The enabling of scientific experiments that are embarrassingly parallel, long running and data-intensive into a cloud-based execution environment is a desirable, though complex undertaking for many researchers.
The
management of such virtual environments is cumbersome and
not necessarily within the core skill set for scientists and engineers.

We present here {\em Chiminey}, a software
platform that enables researchers to
$(i)$ run applications on
both traditional high-performance computing and cloud-based
computing infrastructures,
$(ii)$ handle failure during execution,
$(iii)$ curate and visualise execution outputs,
$(iv)$ share such data with
collaborators or the public, and
$(v)$ search for publicly available
data.

Demo video: http://youtu.be/Twi-d2WT94A
\end{abstract}

\IEEEpeerreviewmaketitle

\section{Introduction}\label{sec:introduction}
Researchers have been using computing resources such as desktops and supercomputers for running their experiments.
In order to use such resources, researchers are expected to know how to set up their execution environment, run their experiments and collect and optionally share the output of their experiments. When executing computational experiments on a local desktop machine, performing these tasks may not be  challenging. However, if an experiment's resource requirements exceed those of a single workstation, then computing environments such as cluster, grid or cloud can be considered.

Cloud computing~\cite{Armbrust2010} presents a unique opportunity for users: it enables researchers to  acquire very large numbers of  computing and storage resources quickly.   
 Moreover, researchers with relatively modest requirements for parallelisation of existing code may be able to avoid learning high-performance computing (HPC) infrastructure concepts. 
Researchers still need to
 learn how to work within a cloud-based environment, which itself presents its own challenges. They need to to create and set up  
  virtual machines (VMs) in the cloud, collect the results of their experiments, and then release the VM resources.
 Furthermore, cloud-based environments are more prone to failure than HPC environments due to network and third-party software issues~\cite{Yusuf2013}, and these environments expect researchers to handle such failures themselves. 

The rate of technological change and innovation for compute environments is ever increasing. 
When a new technology is introduced, both  opportunities and challenges are presented.   
As a researcher migrates from desktop %
to cloud computing, new computing capabilities may be realised; but new skills are required: 
 there needs include not only operational but also fault tolerance and recovery skills. Such challenges distract the  user from focusing on their core goals such as research discovery through creating domain-specific software.

In this paper, we present {\em Chiminey} platform, designed to enable the user to focus on their domain of investigation, and to delegate the platform to deal with the detail that comes with accessing high-performance and cloud computing infrastructure, as well as the data management challenges it poses. 
Researchers are not expected to have a deep technical understanding of cloud-computing, HPC, fault tolerance, or data management in order to leverage the benefits provided by Chiminey. Users may interact with Chiminey via  a web-based graphical user interface or a scriptable API.

We have conducted a number of case studies, applied Chiminey across to two research disciplines in order to assess its practicality:
physics 
and structural biology.  
 The domain  experts appreciated Chiminey's features  
 and noted the time savings for computing and data management.
We believe that Chiminey will have a strong positive impact on the research community, because it gives an opportunity to focus on the main research problems and takes upon itself solving of the major part of the infrastructure problems.

\section{Chiminey}\label{sec:chiminey} 
\begin{figure}
\centering
\scalebox{0.4}{
\includegraphics{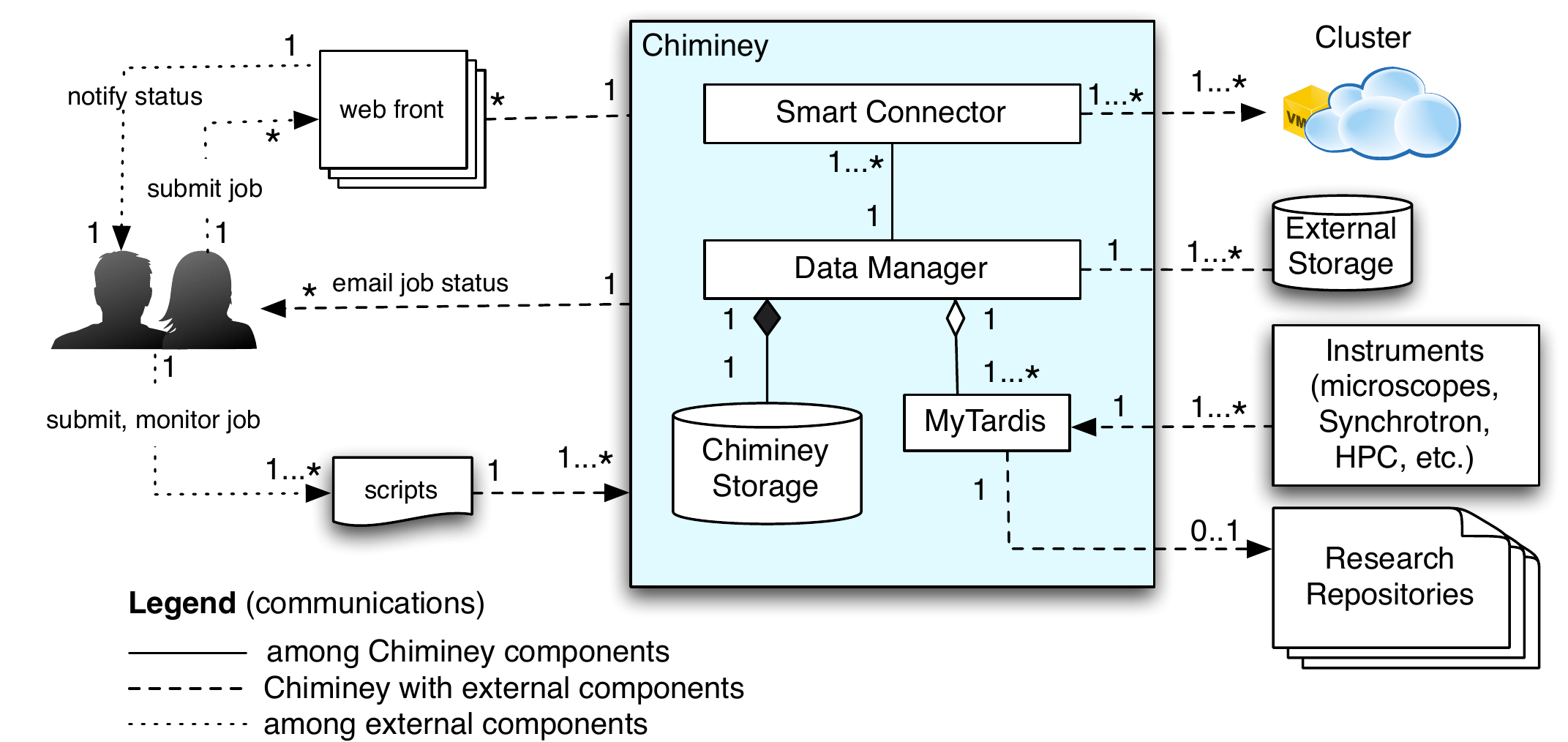}
}
\caption{
UML-based reference architecture of Chiminey platform
}
\label{fig:bdp_reference_arch}
\end{figure}

The {\em Chiminey} platform  
 (cf. Fig.~\ref{fig:bdp_reference_arch}) is a computing and a data management platform that enables researchers
 to perform complex computation  on cloud-based and  
 HPC facilities, handle failure during the execution of their application, curate and visualise execution outputs, share such  data with collaborators or the public, and search for publicly available data.
Chiminey provides a data management platform
both as a source and sink of data coming from instruments and being processed through Chiminey, and as a curation and storage repository for data to utilised by future tools, published and then cited.
 Whenever  HPC computation is completed, its output is transferred to  user-designated locations including a data curator.
 For curating data, Chiminey  uses
{\em MyTardis}~\cite{Androulakis2008}, an application for cataloguing, managing and assisting the sharing of large scientific datasets privately and securely over the web.
MyTardis is currently deployed at various labs and institutions around Australia to capture, manage and provide access to data in research areas such as x-ray crystallography~\cite{Meyer:dz5337}, microscopy, medical imaging, genomics, and HPC.  
The Chiminey platform was created as part of the Bioscience Data
Platform project~\cite{nectar2013}, which is an agile software collaboration
between software engineering and natural sciences researchers. %
Python was chosen as the development language due to its rapid prototyping
features, integration with the MyTardis data curation system, and due
to its increasing uptake by researchers as a scientific software
development language. 
However, the domain-specific calculations could be written in any language (the choice depends on the domain and the concrete research task). %

\emph{Software engineering challenges:}
One of the main challenges is the development of
{\em Smart Connector} (SC) components of Chiminey.
An SC is a core Chiminey component that interacts with a cluster or a cloud service on behalf of the user: it sets up the execution environment (creates the VMs, configures then for the upcoming simulation, etc.), runs a computation, and then transfers the output of the computation to the user's desired location.  SCs handle the provision of cloud-based and HPC infrastructures, as well as give special importance to resource access abstraction and fault tolerance.
The user does not   need to know about   how   VMs  are created and destroyed,  how  a simulation  is configured and executed, or how the final output is transferred.
  With respect to the execution environment, the only information that is expected from the user is to specify the number of computing resources she wishes to use, credentials to access those resources, and the location for transferring the output of the computation.
With respect to configuring and executing the simulation, the user may set the value of domain specific parameters. 
 .

SCs vary from each other by the type of computation to be supported and/or the specific  computing infrastructure to be provisioned.
Chiminey provides a set of APIs to create new and customise existing SCs.
The APIs enable research software engineers to focus on developing the variation point of the new connector rather than  access abstraction and/or fault tolerance support. Chiminey allows  
to specify ranges of input parameter values for a given SC, and subsequently   automatically creating and executing multiple instances of the given connector to sweep across ranges of values.  This allows the researcher to quickly explore a parameter space, and the results to then be visualised.

\begin{figure}[ht!]
\centering
\scalebox{0.35}{
\includegraphics{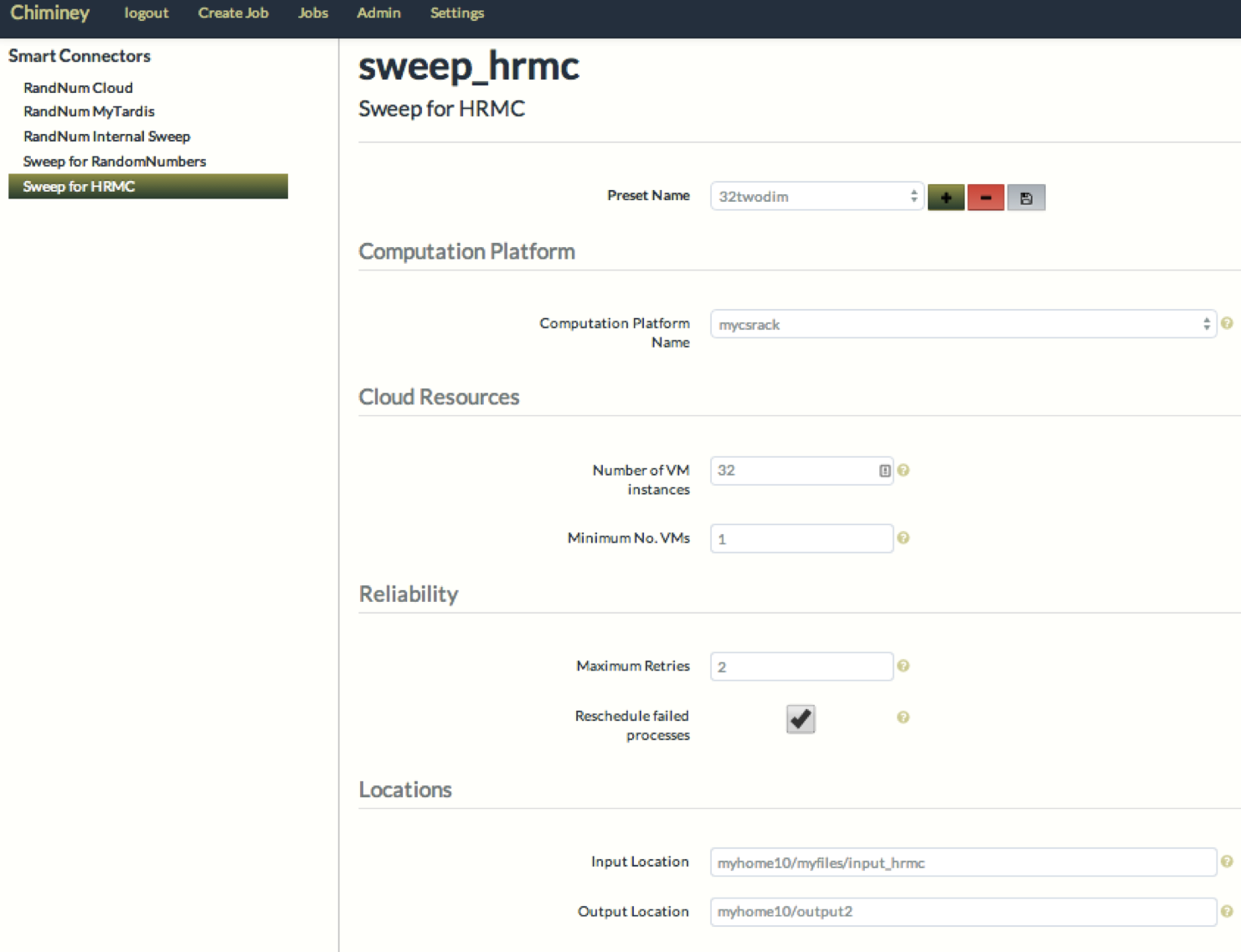}
}
\scalebox{0.35}{
\hspace{32mm}\includegraphics{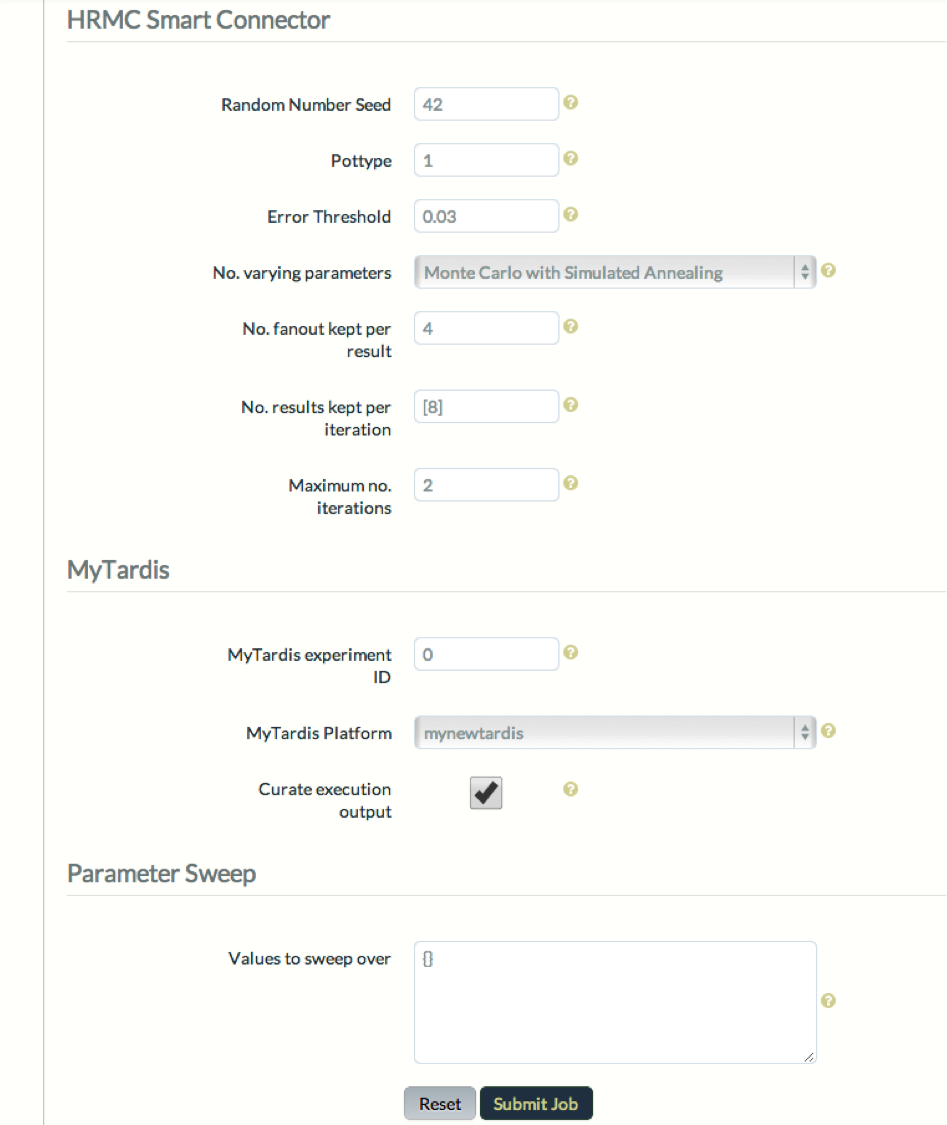}
}
\caption{Chiminey interface:  
specification of an SC execution. The user defines a name of the SC,
chooses the computational platform from a given list,
specifies required cloud resources (desired and minimal number of required VMs),
reliability requirements (maximum number of retries of a failed computation and whether a failed computation should be rescheduled), 
input/output locations, and domain specific characteristics. Finally, the user selects whether the execution output should be curated and where.}
\label{fig:visualisation0}
\end{figure}

\begin{figure}
\centering 
\scalebox{0.4}{
\includegraphics{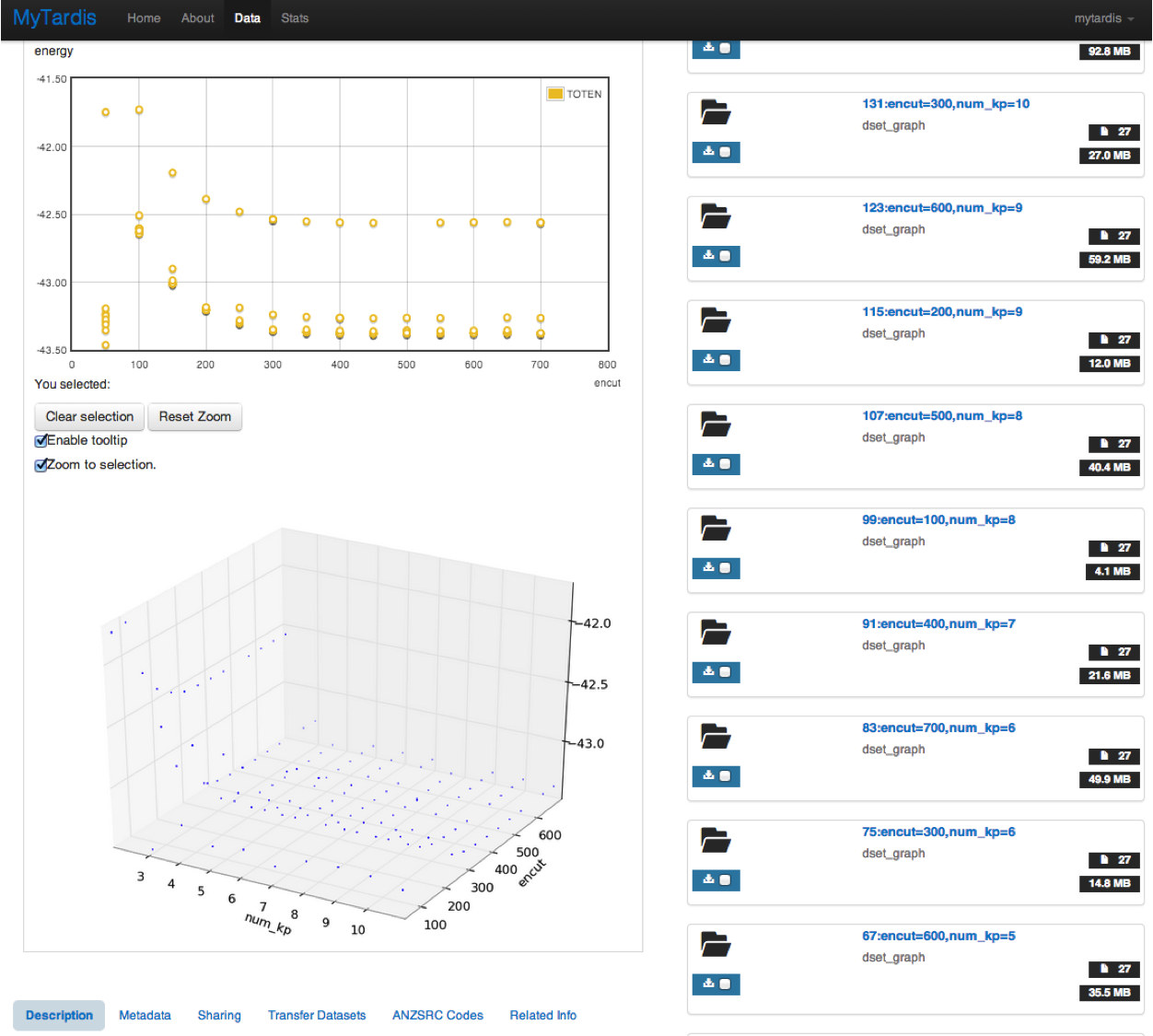}  
}
\caption{Visualisation of computation results in MyTardis, using a plug-in developed  
to provide better usability for the Chiminey platform. The curated datasets are fully accessible and shareable online.
}
\label{fig:visualisation1}
\end{figure}

Another challenge was to develop a user interface which is intuitively clear to the research scientists, who do not have a deep technical understanding of cloud computing, fault tolerance, etc.
The present interface was created on the basis of contextual interviews with the physics
and  biology researchers.  
Fig.~\ref{fig:visualisation0} shows a web interface for the specification of an SC execution.  
Fig.~\ref{fig:visualisation1} shows an example of visualisation:  
two- and three-dimensional graphs are automatically generated  
as physics simulation data is curated.

\emph{Example of a Smart Connector}:
One of the SCs we implemented in Chiminey is a {\em cloud-based iterative MapReduce Smart Connector} (MRSC),
which is suitable for long-running data-parallel programs like Monte Carlo simulations.
These simulations are computational algorithms that rely on repeated random sampling to obtain numerical results:
simulations (so called \emph{MapReduce computations}) should be run many times over, until a predefined criteria are met,
in order to obtain the distribution of an unknown probabilistic entity.
Monte Carlo simulations are often used in solving physical and mathematical problems, especially
for optimization, numerical integration and generation of draws from a probability distribution.

\begin{figure}[ht!]
\centering
\scalebox{0.4}{
\includegraphics{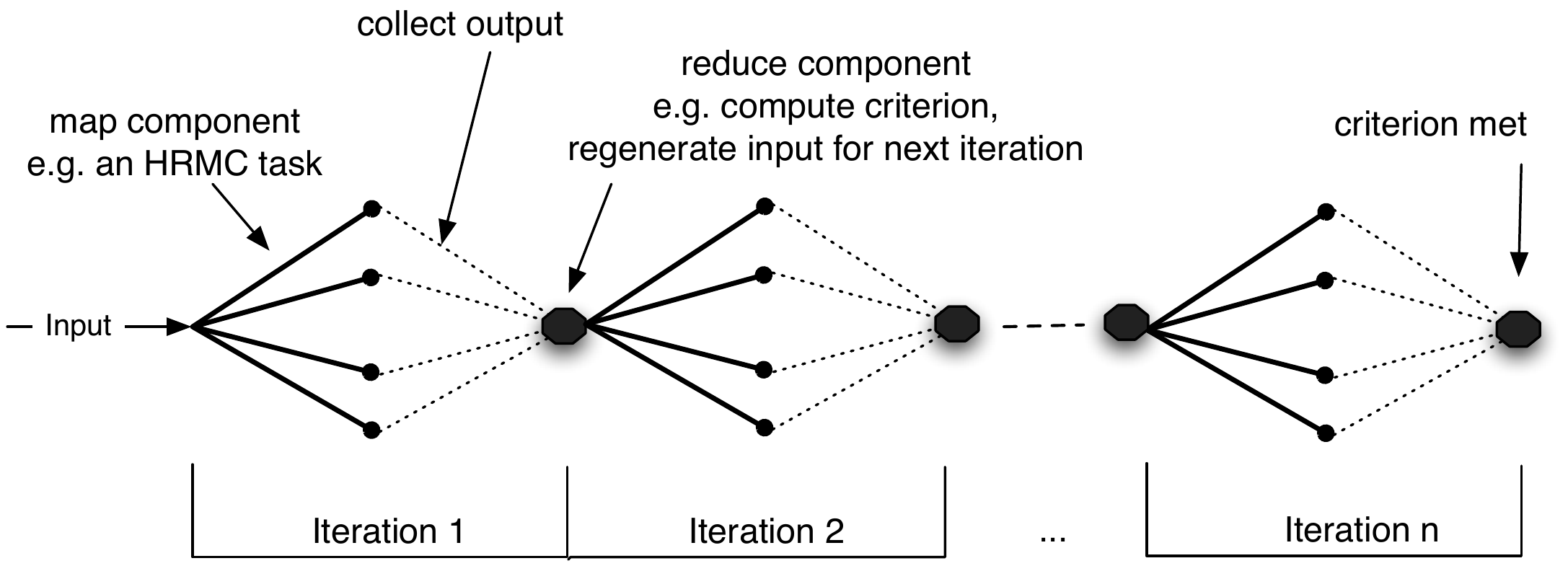} %
}
\caption{The communication and computation pattern of the cloud-based iterative MapReduce Smart Connector}
\label{fig:mr_connector}
\end{figure}

The communication and computation pattern of the MRSC is shown on  Fig.~\ref{fig:mr_connector}.
Since VMs need to be created and configured before execution commences, the execution time of applications should be long enough to justify the use of cloud resources.
The  MapReduce computation is performed iteratively until the predefined criterion is met.
The output of each task is  sent to a MyTardis instance for both generic data access, and domain-specific visualisation. %

\section{Evaluation}\label{sec:evaluation}

We have evaluated our system by running  a number of case studies, involving experiments in physics (material
 characterization) and structural biology (understanding materials at the atomic scale).
  The domain  experts noted the time savings for computing and data management, as well as usability aspects of the
  Chiminey platform.

 The \emph{user interface} allows flexibility in the initial setup, with most
parameters easy to change from their default values for a new exploration of
the parameter space available to the platform.
Finally, the ability to index and store the increased volumes of data
stemming from this new tool are also of high value; academic research is often
conducted under conditions requiring the storage and/or accessibility of data
for several years following the actual work. Chiminey user interface, combined with
the MyTardis data curation module, allows for flexible handling of data
according to its completion and significance. Files can be transferred between
computational resources while the work is in progress, and can be curated when
the workflow slows naturally, such as when the solution to a problem involving
many calculations is found.

In addition, the inclusion of \emph{automated graphing software} means that the user
can easily trace the flow of the calculations through the several sequential
parallel executions that are often required to reach reasonable convergence to
the experimental data. This diagnostic information allows the user to cope with
the increased flow of information available, and judge whether the model is
converging adequately or requires further tweaking. Where the methods can
involve simulated annealing to varied temperatures in an effort to locate the
correct solution, the pathways leading to the best candidate are also of
interest, and the graphs can easily be used for presentations or in written
documents.

In the rest of this section we discuss the application of our tools to execute Monte Carlo simulations.
As these simulations give the basis for
modelling of a material's porosity and the size distribution of its pores, they are
of recent interest in the material
 characterization community.
 One such modeling methodology is the Hybrid Reverse Monte Carlo (HRMC) method~\cite{Opletal2008}.
 This method aims to produce three dimensional atomic coordinates of disordered materials which are consistent with a variety of experimental data
(e.g.,  electron, x-ray and neutron diffraction, porosity information) while ensuring a low energy local bonding environment.
Together with the the  Theoretical Chemical and Quantum Physics group at RMIT University,
we have identified the requirements for a cloud-based execution and deployed this using the MapReduce Smart Connector (cf. Section \ref{sec:chiminey}).  The connector  was configured to
run the HRMC program on the cloud and to manage the output of that execution. The main requirements are (cf. also Fig.~\ref{fig:mr_connector}):
\begin{enumerate}
\item
To exploit
the randomness inherent to the HRMC method for  executing multiple tasks, each with unique input data, in parallel bursts:
this data parallelism  can be satisfied by using the  the {\em map} component of the SC.
\item
To automate a
decision-making process to prune the calculations in accordance with pre-defined
criteria, %
as well as
to regenerate a new
batch of parallel tasks based upon the outcomes of the previous tasks:
this requires computation to be performed on the outputs of all map tasks.
These computations are equivalent to the {\em reduce} component of the MRSC.  
\item
To visualise output of each task to indicate the progress of the calculations with
respect to the  criteria at a glance:
to satisfy the fourth requirement, the output of each task are  sent to a MyTardis instance for both generic data access, and domain-specific visualisation.
\item
To organise and store each output persistently, i.e. to provide data curation: this is achieved by transferring all data to  either  a MyTardis,  
or a user designated location.
\end{enumerate}
In a process reliant on the random nature of Monte Carlo simulation, the
abilities to rapidly process simulation results, make decisions based on
outcomes, and generate new calculations when necessary are of high value.
Here, the automation of the HRMC package via Chiminey has led to significant speedups in
model planning, setup and execution. Where before, the typical modus operandi
was to run a calculation and subsequent evaluation of the pore-size
distribution in the hope of a random match to experimental data, and then to
tweak some initial parameters and try again, now a structured process exists
to facilitate and manage these time-consuming tasks. As well as saving time on
the low-level evaluation and changes to setup, this also has consequent flow-
on effects at the computational level;  restarting calculations no longer have
to wait for user input, meaning that the available resources can be used more
efficiently, with minimal downtime while the problem iterates.

\section{Related Work}\label{sec:relatedwork}
 There are different types of  scientific workflow systems such as Kepler~\cite{Ludascher2006}, Taverna~\cite{Oinn2006} and Galaxy~\cite{Afgan2011}, which are designed to allow researchers to build their own workflows.
 However, Chiminey  provides  drop-in components, i.e.  Smart Connectors,  for  existing workflow engines.
 Researchers utilise and adapt existing Smart Connectors. New types of  Smart Connectors would be developed by the Chiminey development team in collaboration with researchers.
 
  There are  a number of platforms/applications with similar aims and features. In comparison to them, Chiminey provides more features to make the researchers' work more efficient.
  Unlike VIVO, a semantic web application for the discovery of %
  research outputs within an institution \cite{krafft2010enabling},
  the data management component of Chiminey focuses on curating data from \emph{instruments}, %
   visualising and publishing these data, and making the research data itself accessible.
  Unlike Chorus, a web application for managing spectrometry files  \cite{chorus2013}, Chiminey  is not restricted to managing   a specific type of files:
  Chiminey not only  manages any type of files but also allows
 the addition of filters to the files for automatic generation of domain-specific metadata.
 Furthermore, Chiminey provisions a reliable computing capability for data processing.
  Unlike ReDBox, a software platform  for curating and publishing experimental results  \cite{redbox2013},
  Chiminey curates and  publishes metadata and data collected from instruments.
  Furthermore, Chiminey provides a reliable computing and data visualisation capability.

  Nimrod
~\cite{Buyya2000} is a set of software infrastructure for executing large and complex computational processing across several compute resources at a time.
It is compatible with the scientific workflow system Kepler, s.t. users can set up complex computational workflows and have them executed without having to interface directly with an HPC system.
Incorporation the   Nimrod into Chiminey's architecture for the execution of its Smart Connectors is an ongoing work.

\section{ Conclusion}\label{sec:conclusion}
The nature of many scientific problems today  mandates the use of parallel programming to unlock the power of HPC and big data from advanced instruments.
This has  required researchers   to learn HPC, cloud computing and data management skills to address their problems.

We have presented the Chiminey platform, which provides
a reliable computing and data management, and be used by researchers without having to learn  extensive infrastructure concepts and technologies. %
Researchers can  access HPC, use  cloud services,  and  archive, visualise and publish the result of their computations.
In the demo, we have discussed one of our case studies: Monte Carlo simulations. The domain  experts appraised Chiminey with these scenarios, and noted the time savings for computing and data management.

\section*{Acknowledgement}
The Bioscience Data Platform project  
acknowledges funding from the NeCTAR project No. 2179~\cite{nectar2013}.

\balance
\bibliographystyle{abbrv}

\end{document}